\documentclass[preprint2]{aastex6}

\shorttitle{YSGs and RSGs in Cen A} 

\shortauthors{Markakis et al.}

\received{27 Jun. 2017}

\revised{09 Oct. 2017}

\accepted{10 Oct. 2017}

\begin{document}
\title{High resolution observations of Cen A: Yellow and red supergiants in a region of jet-induced star formation? $^{*}$}\thanks{$*$ Based on data collected at Subaru Telescope, which is operated by the National Astronomical Observatory of Japan.}

\author{Markakis, K.\altaffilmark{1,2,5}, Eckart, A.\altaffilmark{1,2}, Castro, N.\altaffilmark{3}, S\'anchez-Monge, \'A.\altaffilmark{1}, Labadie, L.\altaffilmark{1}, Nishiyama, S.\altaffilmark{4}, Britzen, S.\altaffilmark{2}}
\and 
\author{Zensus,  J. A.\altaffilmark{2,1}}

\altaffiltext{1}{I. Physikalisches Institut, Universit\"at zu K\"oln, Z\"ulpicher Str. 77, 50937 K\"oln, Germany}
\altaffiltext{2}{Max-Planck-Institut f\"ur Radioastronomie, Auf dem H\"ugel 69, 53121 Bonn, Germany}
\altaffiltext{3}{University of Michigan, Department of Astronomy, 1085 S. University Avenue, Ann Arbor, MI 48109-1107, USA}
\altaffiltext{4}{Miyagi University of Education, Sendai, Miyagi 980-0845, Japan}
\altaffiltext{5}{markakis@ph1.uni-koeln.de}

\begin{abstract}

We present the analysis of near infrared (NIR), adaptive optics (AO) Subaru and archived HST imaging data of a region near the northern middle lobe (NML) of the Centaurus A (Cen A) jet, at a distance of $\sim15$ kpc north-east (NE) from the center of NGC5128. Low-pass filtering of the NIR images reveals strong -- $>3\sigma$ above the background mean -- signal at the expected position of the brightest star in the equivalent HST field. Statistical analysis of the NIR background noise suggests that the probability to observe $>3\sigma$ signal at the same position, in three independent measurements due to stochastic background fluctuations alone is negligible ($\leq10^{-7}\%$) and, therefore, that this signal should reflect the detection of the NIR counterparts of the brightest HST star. An extensive photometric analysis of this star yields $V-I$, visual-NIR, and NIR colors expected from a yellow supergiant (YSG) with an estimated age $\sim10^{+4}_{-3}$ Myr. Furthermore, the second and third brighter HST stars are, likely, also supergiants in Cen A, with estimated ages $\sim16^{+6}_{-3}$ Myr and $\sim25^{+15}_{-9}$ Myr, respectively. The ages of these three supergiants are in good agreement with the ages of the young massive stars that were previously found in the vicinity and are thought to have formed during the later phases of the jet-HI cloud interaction that appears to drive the star formation (SF) in the region for the past $\sim100$ Myr. \\ \\
\end{abstract}


\section{Introduction}
\label{sec:intro}
NGC5128 is located at a distance of $\sim3.8$ Mpc \citep{2004A&A...413..903R} and is the host of the powerful extra-galactic radio source Centaurus A (Cen A). This peculiar giant elliptical galaxy is considered to be an example of a post-merging system, believed to have formed via the merging of a massive early-type and a smaller gas-rich galaxy \citep[e.g.][]{1983ApJ...272L...5M,1998A&ARv...8..237I}.

At $\sim15$ kpc north-east (NE) from NGC5128's nucleus lies the northern middle lobe (NML) of Cen A. An HI cloud \citep{1994ApJ...423L.101S,2005A&A...429..469O}, CO molecular gas \citep{2000A&A...356L...1C}, shells of old stars \citep{1983ApJ...272L...5M}, filaments of ionized gas \citep{1991MNRAS.249...91M}, young massive stars and OB associations \citep[e.g.][]{1975ApJ...198L..63B,1981ApJ...247..813G,1998ApJ...502..245G,2000ApJ...538..594F,2000ApJ...536..266M,2001A&A...379..781R} have been previously identified near this position. All these are superimposed on a background sheet of older stars, typical of the halo of NGC5128 \citep{2000ApJ...536..266M,2001A&A...379..781R}.

\begin{figure}[t]
\centering
\includegraphics[width=\columnwidth]{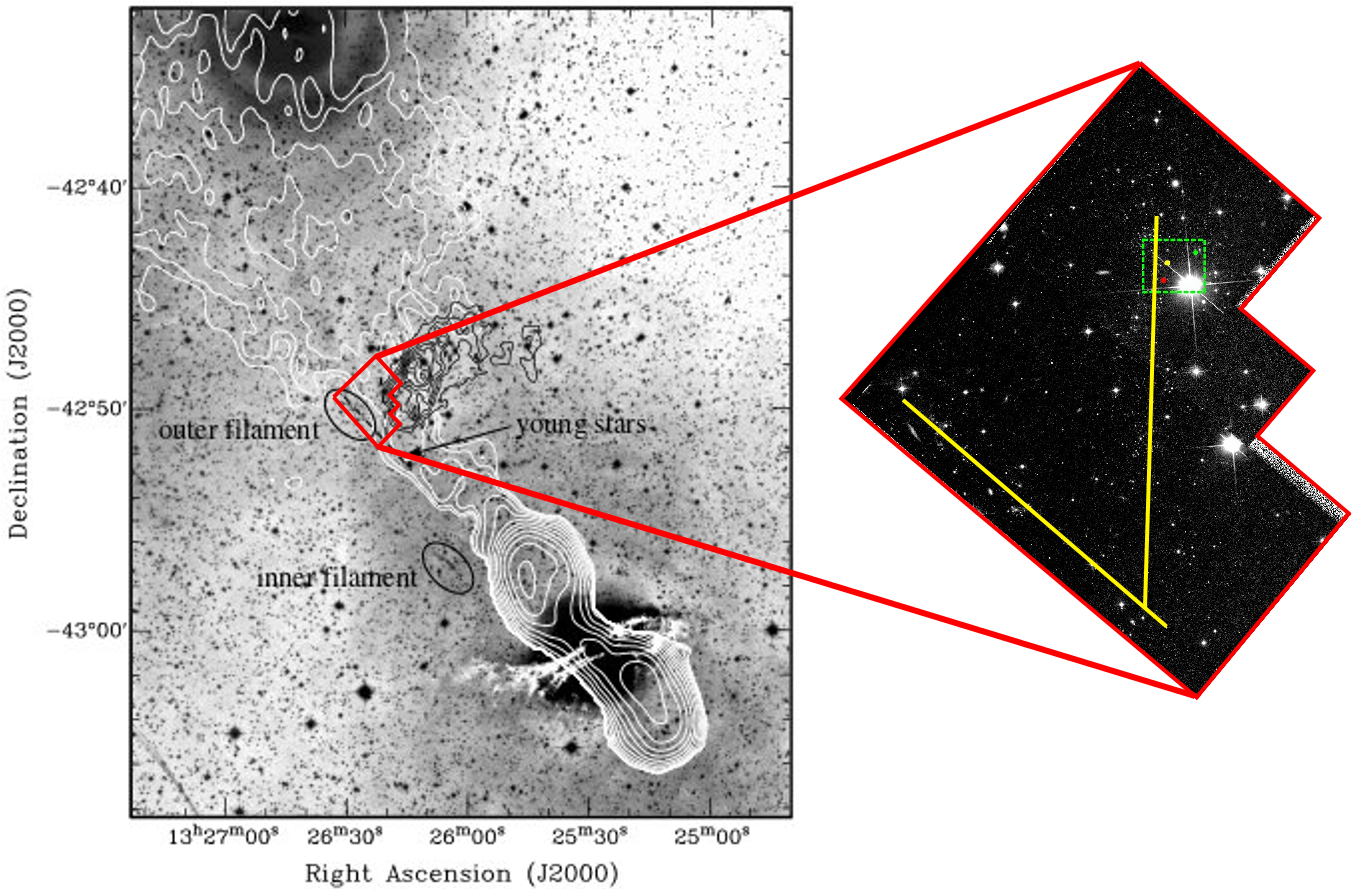}
\caption{Left: Reproduction of Fig. 1 from \citet{2005A&A...429..469O}. The centimeter radio jet (white contours), the HI cloud (black contours), and the filaments of ionized gas (black ellipses) are over-plotted on an optical image of NGC5128. The red outlined region indicates the approximate position and coverage of the HST field (right) studied by \citet{2000ApJ...536..266M}. Right: The yellow lines on the HST $F555W$ image indicate the approximate positions of the young massive stars, while the green dashed rectangle indicates the Subaru field.}
\label{fig:oster}
\end{figure}

Many authors \citep[e.g.][]{1998ApJ...502..245G,2000ApJ...536..266M,2001A&A...379..781R} suggest that the presence of young massive stars in the region is the result of a recent jet-induced star forming episode, since both the gas clouds and the filaments of ionized gas appear to be associated with the young massive stars that lie along the NE edge of the HI cloud and into the jet's path (Fig. \ref{fig:oster}). \cite{2005A&A...429..469O} see anomalous velocities in the south-eastern (SE) tip of the HI cloud and they suggest that this is where the jet-HI cloud interaction takes place, while \cite{2016A&A...595A..65S} find that the gas in the outer filament is likely excited by energy provided by either AGN/shocks, star formation (SF), or a combination of these mechanisms.

The ages of the young massive stars of this region are estimated to be $\sim10-15$ Myr, based on comparisons of their optical color-magnitude diagram (CMD) with both stellar evolutionary isochrones \citep{2000ApJ...538..594F,2001A&A...379..781R} and CMDs of young clusters in the Large Magellanic Cloud \citep{2000ApJ...536..266M}. \cite{2004A&A...415..915R} simulated the recent SF history (SFH) of the region by constructing synthetic CMDs for metallicities $Z=0.004$ and $Z=0.008$, assuming different SF durations and initial mass functions (IMF). A comparison of the synthetic CMDs and the blue main sequence (MS) luminosity functions with the ones observed by \cite{2001A&A...379..781R} revealed that the SF in the halo of NGC5128 is, likely, not an episodic event, but a continuous process for (at least) the last $\sim100$ Myr, which has either ceased $\sim2.5$ Myr ago, or is still ongoing. They also constrained the IMF slope $\alpha \sim 2-2.6$ and the SF rate $\sim 0.004-0.013 M_{\sun} yr^{-1}$. The simulations of \cite{2004A&A...415..915R} appear to be consistent with recent findings of \cite{2016A&A...586A..45S}, who estimate the metallicity and the SF rate in the outer filament to be $<Z> \sim 0.6Z_{\sun} = 0.008$ and $\sim 0.004 M_{\sun} yr^{-1}$, respectively.

Within the suggested time-frames, many of these young massive stars are expected to have evolved past the MS. This makes the NE part of NGC5128 an excellent target for searching for extra-galactic NIR-excess sources, e.g. later-type-- yellow (YSG) and red (RSG)-- supergiants and asymptotic giant branch (AGB) stars, that should exist within the framework of a recent (jet-induced) SF. Attempting, therefore, to observe a denser-than-average region outside our Local Group (LG) in the NIR, using a state-of-the-art 8-m class telescope equipped with an adaptive optics (AO) system, is expected to reveal a wealth of information regarding the nature of its evolved massive stellar populations and, potentially, also of the history of the jet-HI cloud interaction itself.

This paper is organized as follows: In Sect. \ref{sec:observations} we describe the data and photometric calibration; in Sect. \ref{sec:FoV_locate} we describe the process of locating the Subaru field on archived HST images of the same region; in Sect. \ref{sec:data_processing} we describe the near infrared (NIR) data processing and we attempt, both statistically and photometrically, to constrain the spectral types and luminosity classes of the three brighter stars in the HST field; and in Sect. \ref{sec:discussion} we discuss our results and the observational potential of using AO assisted observations in galaxies outside our LG.

\section{Observations and data}
\label{sec:observations}

\subsection{Subaru data}
\label{SUB_data}
The data-set used for studying the young stellar population along the jet of NGC5128 consists of AO assisted NIR data in the $J$($10$), $H$($10$), and $K_{S}$ ($10$) bands (number of images), taken on 2012 May 17, with the Subaru telescope at Mauna Kea, Hawaii, using the HiCIAO \citep{2010SPIE.7735E..30S} instrument. The exposure time is $t_{exp}^{J,H,K_{S}} = 60$ sec for each individual frame in all bands. The AO188 AO system is used \citep{2010SPIE.7736E..0NH}. It is equipped with a 188-element wavefront curvature sensor with photon counting APD modules and a 188 element bimorph mirror, installed at the IR Nasmyth platform of the Subaru telescope. As a result, the angular resolution of the data is $\sim270$, $\sim210$, and $\sim170$ mas for the $J$, $H$, and $K_{S}$ bands, respectively. The $2048\times 2048$ pixel$^2$ Hawaii-IIRG HgCdTe detector provides a pixel scale of $0.010$ $arcsec~pixel^{-1}$ and a field of view (FoV) of $20 \times 20$ arcsec$^2$. The studied field was chosen because of the presence of bright foreground natural guide stars, which is a prerequisite for the AO system to deliver the highest angular resolution possible.

No reduction package was available for HiCIAO, so a pipeline was developed from scratch in order to correct for the high frequency 32-strip artifact noise, introduced by the 32 readout channels of the detector. All images were dithered and have undergone the usual bad pixel correction, flat-fielding (dome-flat), sky-subtraction, alignment, and median stacking. Finally, we remove any large-scale patterns from our images by subtracting an image of the weighted average of the mean row and column values, respectively. In all images north is up and east is left. The reduced ready-for-science $\sim20\times18$ arcsec$^2$ $J$, $H$, and $K_{S}$ images are shown in Fig. \ref{fig:our_field}.

\subsubsection{Subaru photometric calibration}
\label{calib}

The flux calibration of the Subaru data was performed using standard star observations in our data-set to derive appropriate zero points for the $J$,$H$, and $K_{S}$ bands. Apparent magnitudes are then calculated using:

\begin{equation}
m^i_{Sub.} = \mathrm{Z.P.^i_{Sub.}} - 2.5 \log({\mathrm{Counts^i_{Sub.}} \over \mathrm{Exp.Time^i_{Sub.}}})~~.
\end{equation}

\begin{figure}[t]
\centering
\includegraphics[width=4cm]{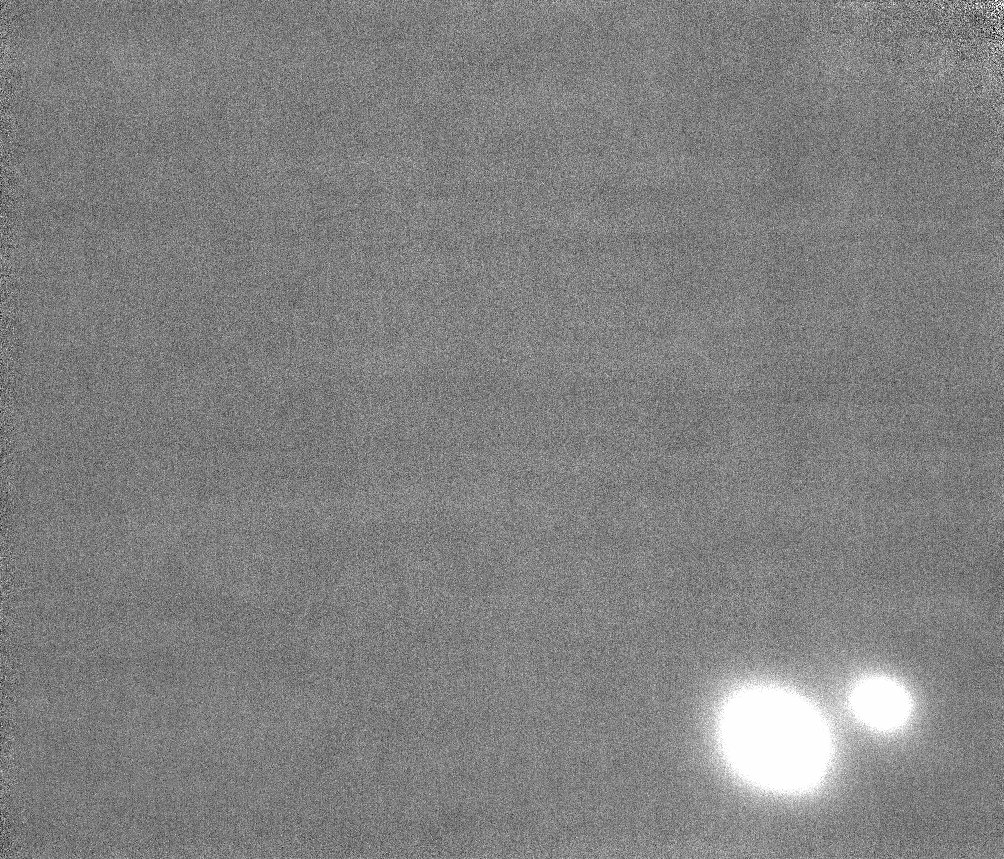}
\includegraphics[width=4cm]{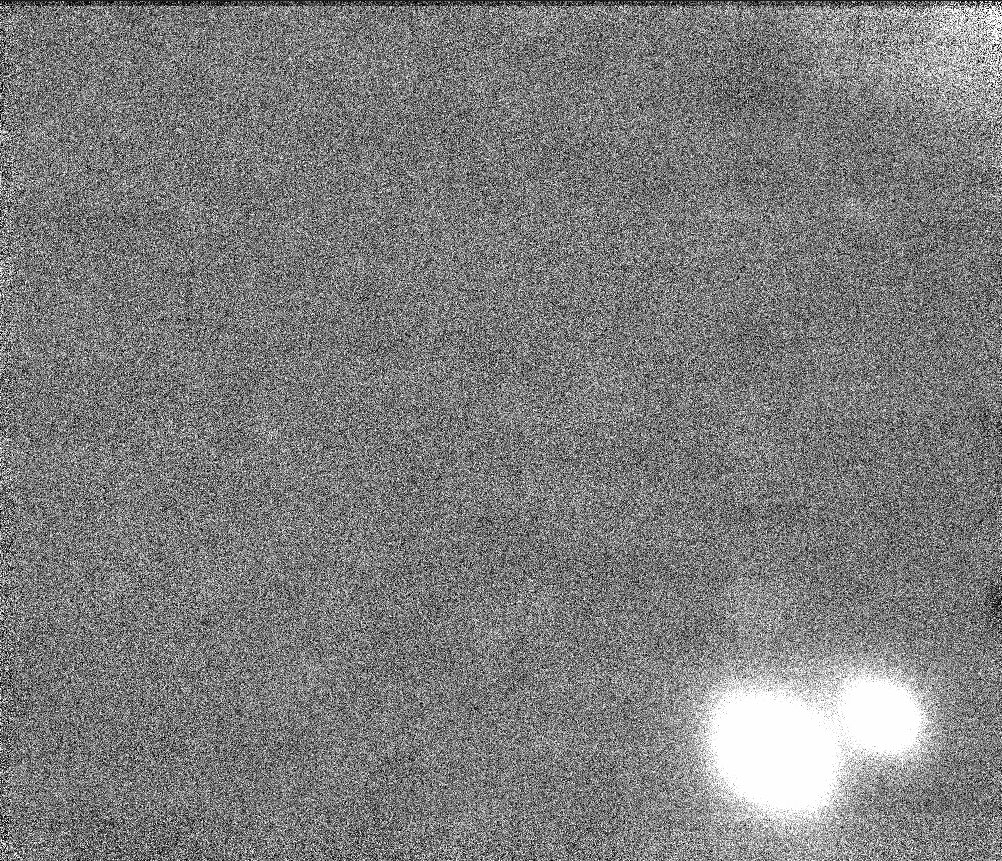}
\includegraphics[width=4cm]{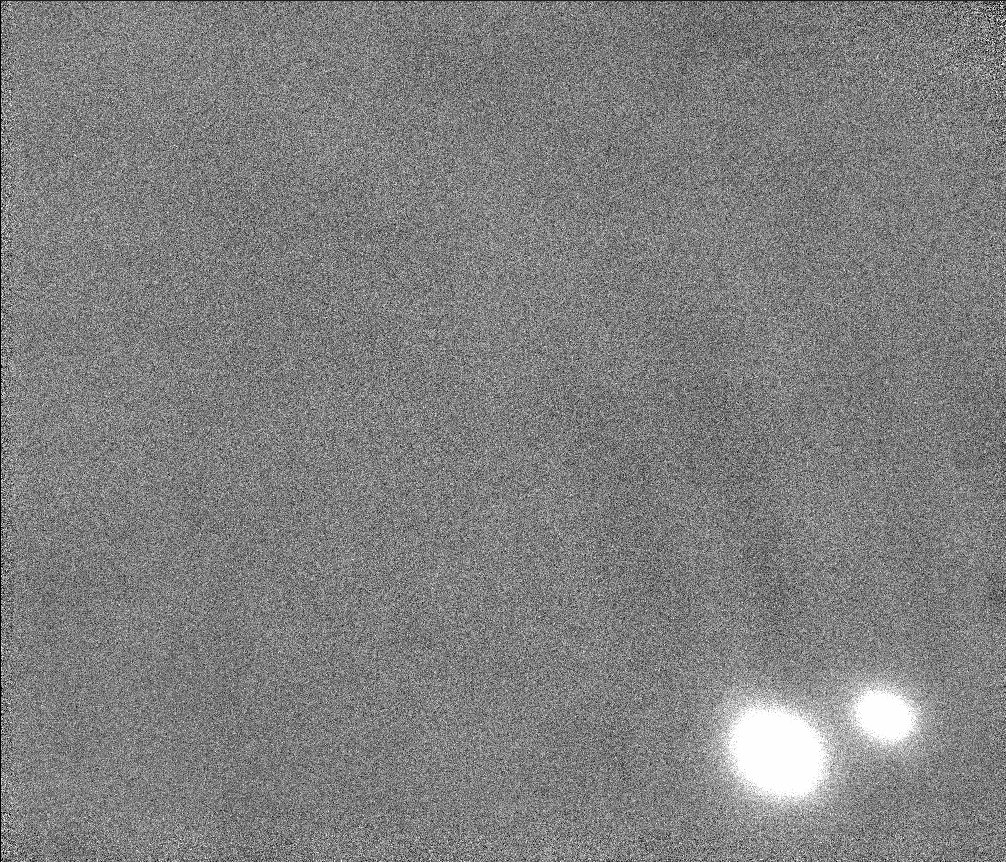}
\caption{Subaru $J$(top left), $H$(top right) and $K_{S}$(bottom middle) bands, $\sim20\times18$ arcsec$^2$ ready-for-science images.}
\label{fig:our_field}
\end{figure}

The two bright (AO guide) stars\footnote{Hereafter ``guide stars''.} in the bottom right corner of Fig. \ref{fig:our_field} can be used to test the quality of the photometric calibration against 2MASS photometry of the same sources. Due to the lower angular resolution of the 2MASS images, however, these sources appear as an unresolved point source\footnote{Hereafter G.S.} with $m_{2MASS~G.S.}^{K_{S}} = 11.15 \pm 0.02$ mag, $m_{2MASS~G.S.}^{H} = 11.30 \pm 0.04$ mag, and $m_{2MASS~G.S.}^{J} = 11.51 \pm 0.03$ mag. The equivalent G.S. Subaru magnitudes are given by the sum of the fluxes of the SE (G.S.1) and the north-western (NW -- G.S.2) guide stars, namely $m_{SUB.~G.S.}^{K_{S}} = 11.17 \pm 0.14$ mag, $m_{SUB.~G.S.}^{H} = 11.33 \pm 0.14$ mag, and $m_{SUB.~G.S.}^{J} = 11.56 \pm 0.14$ mag. The very good agreement between 2MASS and Subaru magnitudes indicates that our calibration is fairly accurate, which allows us to use the two NIR guide stars as calibration reference sources for the rest of our magnitude estimations.

\subsection{HST data}
\label{HST_data}
We also use archived HST $F555W$ and $F814W$ images of the same region, originally analyzed by \citet{2000ApJ...536..266M} (Fig. \ref{fig:SUB_HST_FOV}). We obtain $F555W$ and $F814W$ magnitudes for the stars in our interest, by performing aperture photometry. Transformations between the STMAG and the Johnson UBVRI photometric systems are subsequently performed according to the tables of the WFPC2 Photometry Cookbook/HST Data Handbook for WFPC2. Throughout this paper we adopt average $V-F555W=-0.01\pm0.06$ mag and $I-F814W=-1.30\pm0.24$ mag. We test the validity of the adopted transformations against the V band magnitudes and $V-I$ colors of the two brightest stars on the WF2 chip \citep[see Table 1,][]{2000ApJ...536..266M}. Our measurements are within $\sim5\%$ with respect to the published values, indicating that the chosen transformations are fairly accurate.

\section{Locating the Subaru field}
\label{sec:FoV_locate}

\begin{figure}[t]
\centering
\includegraphics[width=4cm]{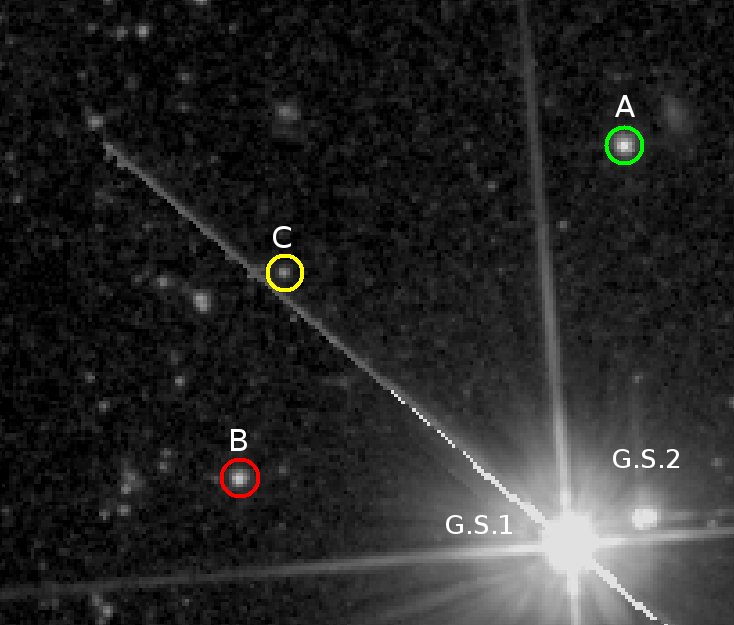}
\includegraphics[width=4cm]{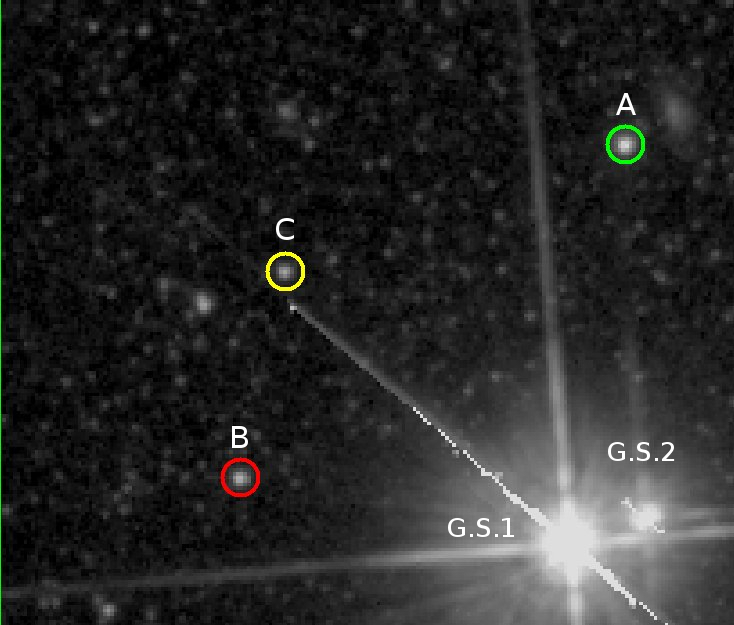}
\caption{HST $F555W$ (left) and $F814W$ (right) images. This is the equivalent to the Subaru $J$, $H$, and $K_{S}$ field ($\sim20\times18$ arcsec$^2$). The colored circles mark the positions of the three brighter stars on the $F814W$ HST field. The A HST star is the brightest, the B is the second brighter, and the C is the third brighter.}
\label{fig:SUB_HST_FOV}
\end{figure}

The Subaru field is located at $\sim15$ kpc NE from the center of NGC5128, with central coordinates $\alpha_{2000}$ = 13$^h$26$^m$21$^s$.127 and $\delta_{2000}$ = -42$^o$49$'$09$''$.02. In order to precisely locate it on the HST images (Fig. \ref{fig:SUB_HST_FOV}), we use the two guide stars and the cardinal directions as reference points. 

The projected separations and position angles (PA) of the guide stars are $2.19\pm0.14$ and $2.25\pm0.01$ arcsec and $19.3\pm1.0$ and $19.0\pm0.1$ degrees on the HST and Subaru images\footnote{The centers of both Subaru guide stars and of the fainter (NW) HST guide star are estimated by Gaussian fitting on the point-spread-function (PSF). For the brighter (SE) HST guide star, however, we use the central pixel of the saturated peak of the PSF as an acceptable center estimation, but with a significantly larger uncertainty.}, respectively. These indicate that the guide stars are either an apparent projected, or a long period physical and not a close physical binary system. The insignificant change in their separation and PA between the epochs of the HST and the Subaru data (i.e. $\sim15$ yrs) could be the result of either a ``lucky'' snapshot of a close physical binary in the same orbital configuration as on the HST images, or the combination of the different (yet negligible) proper motions of the two projected members. The latter seems more likely in a, time-wise, random observation. 

Moreover, the total (i.e. over 15 yrs) G.S. proper motion of $24.2\pm14.3$ mas in RA and $-34.5\pm12.0$ mas in Dec \citep[UCAC4 catalog, ][]{2013AJ....145...44Z} can, in principal, account for the small changes in the separation and PA of the guide stars. Most importantly though, the negligible G.S. proper motion indicates that the HST field stars\footnote{The term ``field stars'' refers to the faint background stars seen on the HST images (Fig. \ref{fig:oster} and \ref{fig:SUB_HST_FOV}), which are thought to be part of the young blue stars along the Cen A jet.} are expected to be found at, roughly, the same-- relative to the guide stars-- positions on the Subaru $J$, $H$, and $K_{S}$ images as well.

\section{Data processing}
\label{sec:data_processing}

\begin{figure}[t]
\centering
\includegraphics[width=4cm]{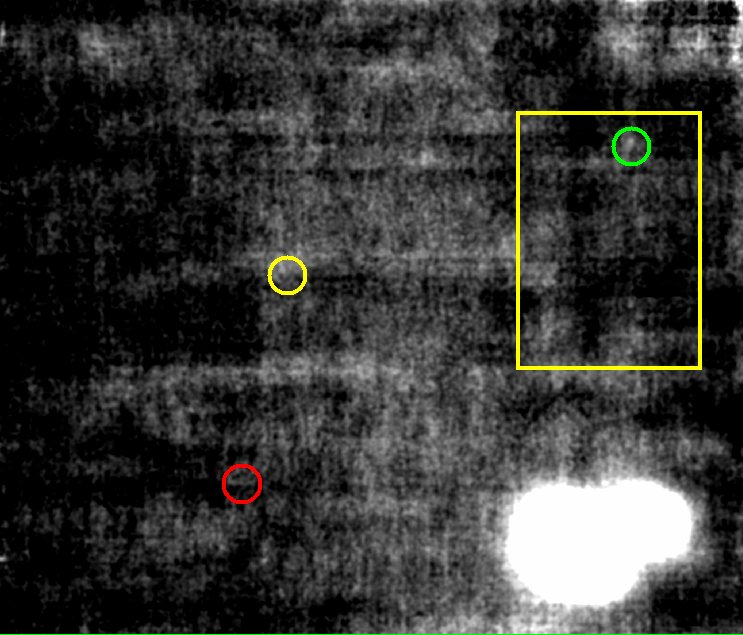}
\includegraphics[width=4cm]{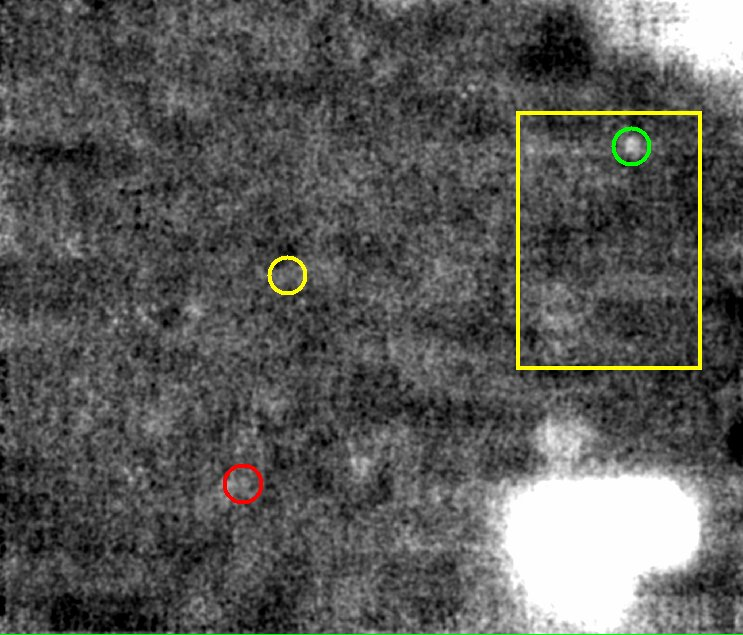}
\includegraphics[width=4cm]{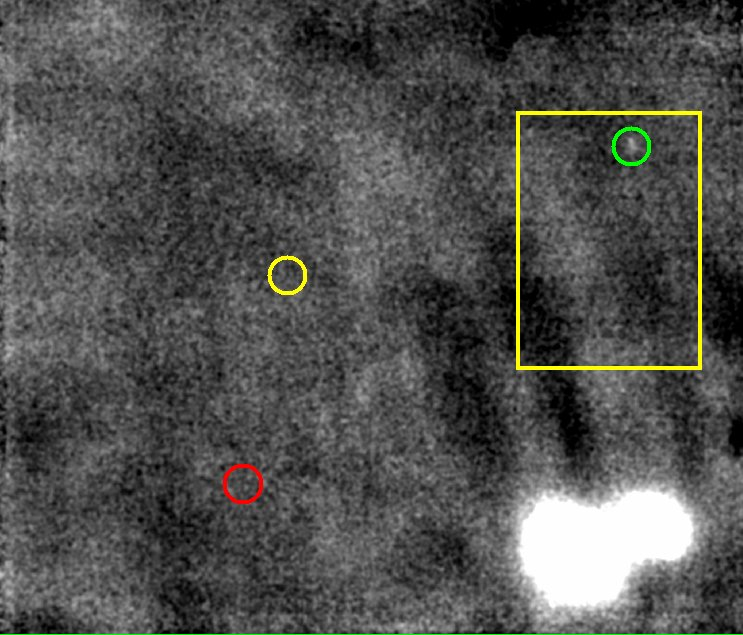}
\includegraphics[width=4cm]{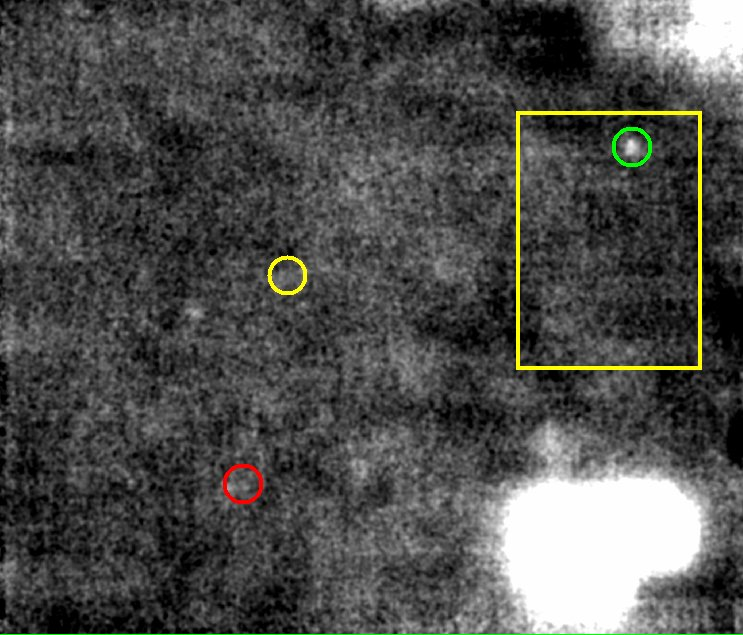}
\caption{Subaru $J$(top left), $H$(top right), $K_{S}$(bottom left), and the $JHK_{S}$ composite (bottom right) images, low-pass filtered with a 10-pixel Gaussian kernel. The three colored circles mark the expected positions of the three brighter stars on the HST $F814W$ field, while the $\sim5\times7$ arcsec$^2$ statistics region is outlined by the yellow box.}
\label{fig:our_field_clipped}
\end{figure}

Initially, there is no obvious signal on the Subaru images (Fig. \ref{fig:our_field}) that can be directly correlated with the HST field stars (Fig. \ref{fig:SUB_HST_FOV}). We apply a low-pass filter with a 10-pixel Gaussian kernel on the NIR data, in order to increase the signal-to-noise-ratio (SNR) and highlight any signal that might be embedded into the noise. A $JHK_{S}$ composite image is also constructed from the low-pass filtered NIR images (using equal weighting for each band), to achieve a further increase of the SNR by a factor of $\sim\sqrt3$ (Fig. \ref{fig:our_field_clipped}).

\begin{figure}[t]
\centering
\includegraphics[width=2.7cm]{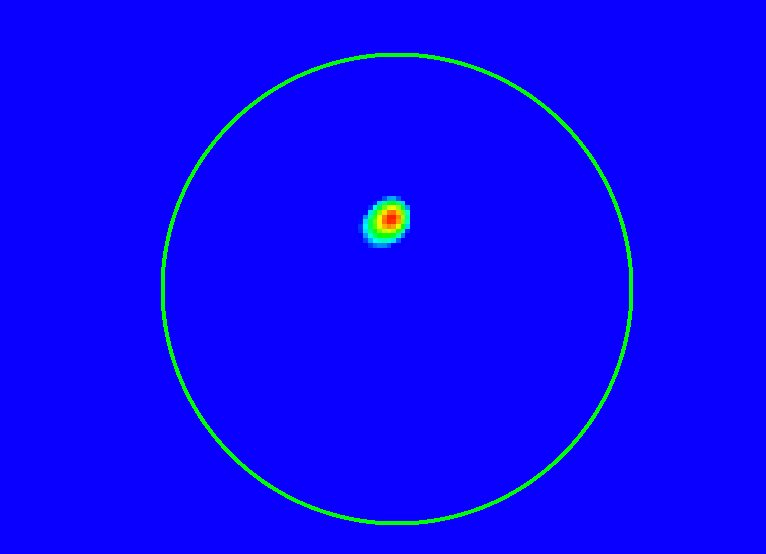}
\includegraphics[width=2.7cm]{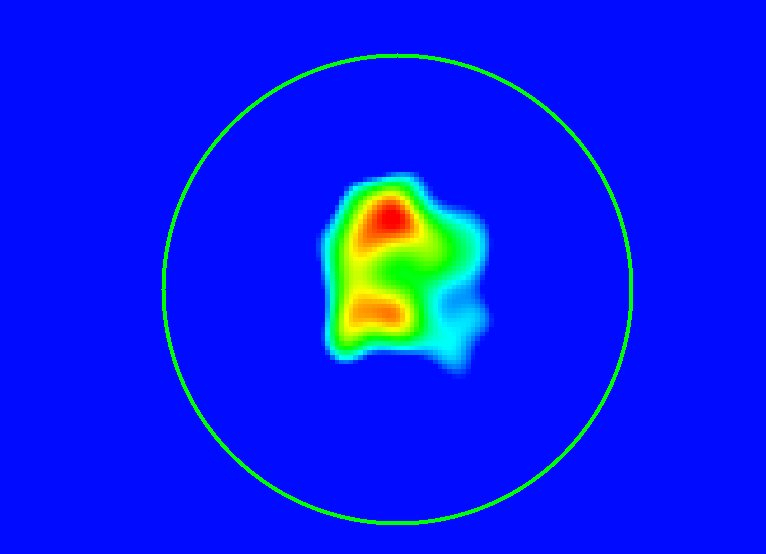}
\includegraphics[width=2.7cm]{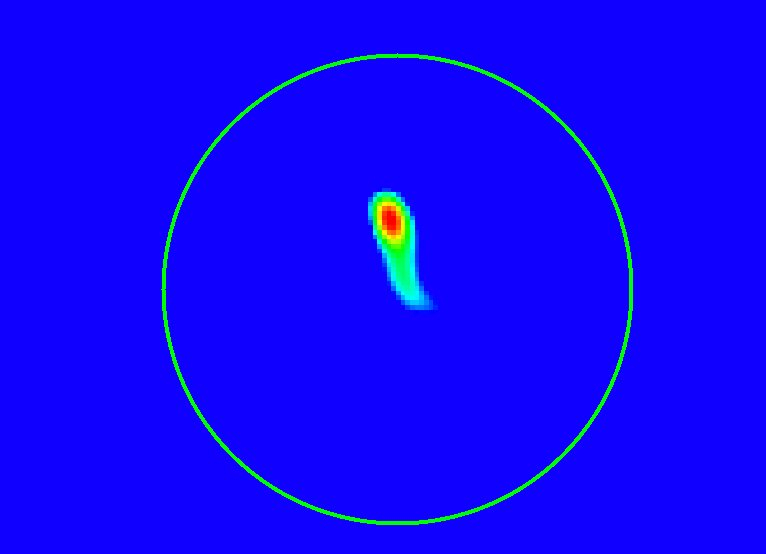}
\includegraphics[width=8.5cm]{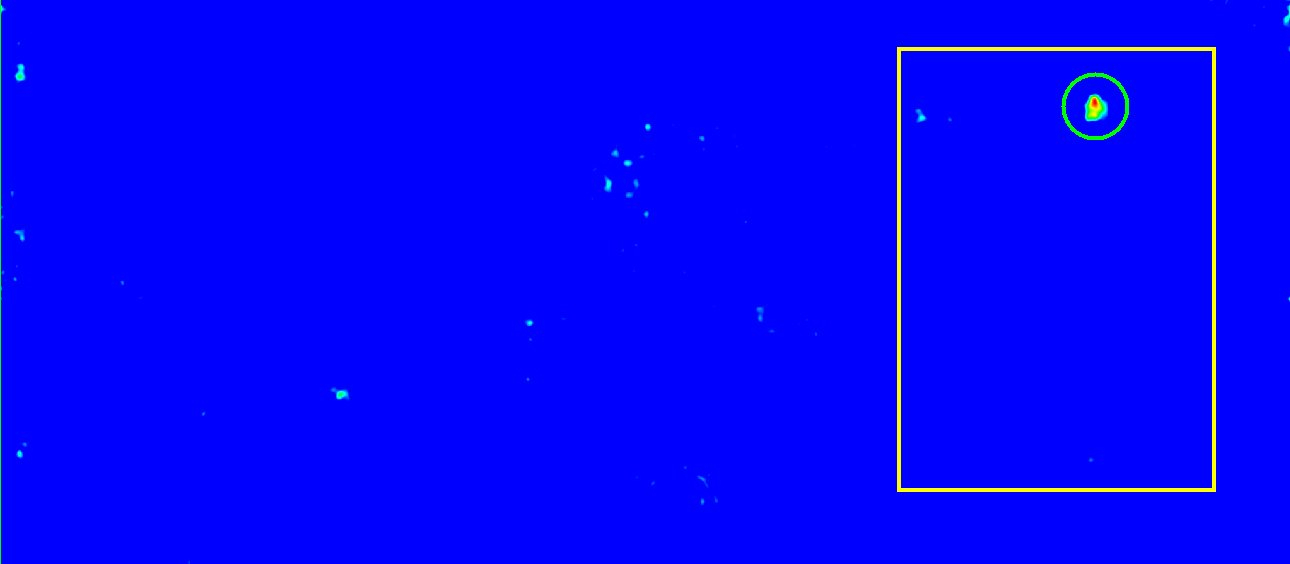}
\caption{Top: The $>3\sigma$ NIR signal at the expected position of the A HST star (zoomed-in region) in the Subaru $J$(left), $H$(middle), and $K_{S}$(right) bands, respectively. Bottom: A $\sim20\times9$ arcsec$^2$ region of the Subaru $JHK_{S}$ composite with the statistics region outlined by the yellow box ($\sim5\times7$ arcsec$^2$). All images are low-pass filtered with a 10-pixel Gaussian Kernel and clipped at $3\sigma$ above the background mean (blue background).}
\label{fig:our_field_statistical}
\end{figure}

All the low-pass filtered images of Fig. \ref{fig:our_field_clipped} show \emph{strong signal} at the expected position of the brightest HST field star (A HST star -- the star in the green circle in Fig. \ref{fig:SUB_HST_FOV}). The peak flux of this signal is $>3\sigma$ above the background mean, while this is the only position where a signal with these characteristics is consistently present in all NIR bands\footnote{For the $H$ band this refers to the northern peak. The southern peak is not present in any other NIR band, which suggests that it could have originated from a low/moderate (e.g. a $\sim+0.5 - 1 \sigma$) random noise fluctuation superimposed on the $>3\sigma$ signal, mimicking a secondary peak feature. Such a small random fluctuation, however, could occur with a rather significant probability of $\sim15-30\%$. Therefore, we do not consider it as a true signal for the rest of this work. We refer the reader to Sect. \ref{statistical_approach} for a more extended discussion on the background noise of an astronomical image.} (Fig. \ref{fig:our_field_statistical}). Since the field stars are expected to be found at the same positions on both the Subaru and the HST images, can the observed $>3\sigma$ signal come from the NIR counterparts of the A HST star and, if yes, can we constrain its nature? Answering this requires both a statistical and a photometric analysis.

\subsection{Statistical analysis of the NIR signal}
\label{statistical_approach}

Approaching the above question statistically means that we have to formulate it in a more mathematical way: What is the probability to observe $>3\sigma$ signal at the same position in three independent experiments, if it results from stochastic background noise fluctuations alone?

Generally, the values of the background noise of an astronomical image follow the Poisson distribution, which for large statistical samples should approach the Gaussian distribution. Given a Gaussian distribution, the probability to randomly draw (observe) a variable (a flux value) $F_{i}$ that satisfies the condition \\

\begin{displaymath}
\bar{F_{bg}}-3\sigma\geq~F_{i}~~\mathrm{and}~~F_{i}\geq\bar{F_{bg}}+3\sigma~~, \\
\end{displaymath}

\noindent
where $\bar{F_{bg}}$ is the mean (mean flux) of the distribution, from a single experiment (observation) is $\leq0.27\%$.

We perform our statistical analysis in a sub-region of $\sim5\times7$ arcsec$^2$ (yellow box in Figs. \ref{fig:our_field_clipped} and  \ref{fig:our_field_statistical}, $N\sim3.5\times10^{3}$ pixels), including the position of the A HST star. Given the large statistical sample, the background noise distribution is expected to be very close to normal. Our measurements within this sampling area support this view, since the mean and the median converge, implying a) that the distribution is not heavy tailed and, more importantly, b) that the chosen statistical sample represents the global statistics well, at least, in terms of statistical power. 

Therefore, if we randomly select a pixel inside our sampling area from one observation, the probability to observe a value $\geq3\sigma$ above the background mean is $\leq0.135\%$ or $P(A)\leq0.00135$\footnote{$P(A)\leq0.0027$ is the probability to observe values outside the $\bar{F_{bg.}}\pm3\sigma$ limits. Half of this should account for the probability to observe values that exceed only the upper limit of $+3\sigma$. Therefore, a $P(A)\leq0.00135$ is adopted as a realistic occurrence probability of a stochastic $\geq3\sigma$ observation in a single experiment.}. Considering also that the $>3\sigma$ signal at the position of the A HST star is the only consistently present signal in all our NIR images (i.e. three independent experiments), then the probability that it results from random background noise fluctuations should be $[P(A)]^{3}\leq2.5\times10^{-9}$ or $\leq2.5\times10^{-7}\%$. Since this is negligible, we can safely consider that the presence of an underlying point source, which systematically ``drives'' the flux within the green circle above the $+3\sigma$ limit in repeated observations, is the origin of the observed $>3\sigma$ NIR signal of Fig. \ref{fig:our_field_statistical}. 

Are, however, the photometric properties of the $>3\sigma$ NIR detections consistent with the visual photometric properties of the A HST star? 

\subsection{Photometric calibration of the NIR detections}
\label{photometric_approach}
The $V$ and $I$ magnitudes of the A HST star are measured with aperture photometry on the $F555W$ and $F814W$ HST images (Sect. \ref{HST_data}). They are $m^{V}_{HST,A} = 20.23 \pm 0.12$ and $m^{I}_{HST,A} = 19.39 \pm 0.26$ mag, respectively.

On the contrary, due to the low SNR of the Subaru data, aperture photometry is not an option for estimating the NIR magnitudes of the $>3\sigma$ detections. Alternatively, their $J,H$, and $K_{S}$ magnitudes are estimated by taking the ratio of the sum of the peak fluxes of G.S.1 and G.S.2 to the peak fluxes of the $>3\sigma$ NIR detections\footnote{Point sources of different brightness observed simultaneously with the same telescope are convolved with the same PSF as long as their separation is smaller than $\sim30$ arcsec (i.e. isoplanatic angle). Consequently, the ratio of the fluxes at any point across their PSFs and the ratio of their integrated fluxes will (ideally) be approximately equal and, therefore, any of these measurements should be representative of their true brightness difference.}, using
\\
\begin{eqnarray}
&m^{J,H,K_{S}}_{>3\sigma} = m^{J,H,K_{S}}_{G.S.} + 2.5 \times\log X\\
&\mathrm{with},~~~~ X = \left({F^{J,H,K_{S}}_{G.S.1} + F^{J,H,K_{S}}_{G.S.2} \over F^{J,H,K_{S}}_{>3\sigma}}\right)~~,\nonumber
\label{eq:residual mag}
\end{eqnarray} 

\noindent
where, $F^{J,H,K_{S}}_{G.S.i}$ is the peak flux of each guide star, $F^{J,H,K_{S}}_{>3\sigma}$ is the peak flux of each $>3\sigma$ detection, and $m^{J,H,K_{S}}_{G.S.}$ is the Subaru $J/H/K_{S}$ magnitude of the G.S. that we derived in Sect. \ref{calib}. 

This yields $m^{K_{S}}_{>3\sigma} = 18.15 \pm 0.37$, $m^{H}_{>3\sigma} = 18.51 \pm 0.36$, and $m^{J}_{>3\sigma} = 18.96 \pm 0.36$ magnitudes for the $>3\sigma$ detections in each NIR band. The uncertainties are calculated according to the relation:

\begin{eqnarray}
&\delta m^{J,H,K_{S}}_{>3\sigma} = \left((\delta m^{J,H,K_{S}}_{G.S.})^2 +0.43^2\times2.5^2\times\left({\delta X \over X}\right)^2\right)^{1/2}~~~~\\
&\mathrm{with}, \left({\delta X \over X}\right)^2 = {(\delta F^{J,H,K_{S}}_{G.S.1})^2 + (\delta F^{J,H,K_{S}}_{G.S.2})^2 \over (F^{J,H,K_{S}}_{G.S.1} + F^{J,H,K_{S}}_{G.S.2})^2} + \left({\delta F^{J,H,K_{S}}_{>3\sigma} \over F^{J,H,K_{S}}_{>3\sigma}}\right)^2 \nonumber\\
&\mathrm{and}~(\delta m^{J,H,K_{S}}_{G.S.})^2=(\delta m^{J,H,K_{S}}_{G.S.1})^2 + (\delta m^{J,H,K_{S}}_{G.S.2})^2 \nonumber
\label{eq:residual uncertainties}
\end{eqnarray} 

\noindent
where we assume a $10\%$ uncertainty in the measurement of $F^{J,H,K_{S}}_{G.S.i}$ and a $30\%$ uncertainty in the measurement of the, more uncertain, $F^{J,H,K_{S}}_{>3\sigma}$.

Having magnitude estimations, we can attempt to constrain the $V-I$, visual-NIR\footnote{i.e. $V-J$, $V-H$, and $V-K_{S}$.}, and NIR\footnote{i.e. $J-H$, $H-K_{S}$, and $J-K_{S}$.} colors of the $>3\sigma$ NIR detections/A HST star and, therefore, its spectral type and luminosity class.

\begin{table}[t]
\caption{NIR and Visual-NIR colors.}
\label{tab:colors}
\begin{center}
\resizebox{\columnwidth}{!}{
\hspace*{-3.2cm}
\begin{tabular}{|c|c|c|c|c|c|c|c|}\hline \hline
Colors   & $V-I$       & $V-J$   & $V-H$      & $V-K_{S}$       & $J-H$ \tablenotemark{a}      & $H-K_{S}$ \tablenotemark{a}    & $J-K_{S}$ \tablenotemark{a}\\\hline

A HST Star & \textcolor{black}{0.84}        & \textcolor{black}{1.27}        & \textcolor{black}{1.72}        & \textcolor{black}{2.08}        & \textcolor{black}{0.45}        & \textcolor{black}{0.36}        & \textcolor{black}{0.81}\\\hline
Uncertainties & ($\pm0.28$) & ($\pm0.38$) & ($\pm0.38$) & ($\pm0.39$) & ($\pm0.51$) & ($\pm0.52$) & ($\pm0.52$)\\\hline\hline
Spectral Type & \multicolumn{7}{c|}{Intrinsic colors of Supergiants \citep{2001ApJ...558..309D}} \\\hline

A5   &  0.180 &  0.200 &  0.290 &  0.350 &  \textcolor{blue}{0.090} &  \textcolor{blue}{0.060} &  0.150\\\hline
F0   &  0.310 &  0.360 &  0.510 &  0.600 &  \textcolor{blue}{0.150} &  \textcolor{blue}{0.090} &  0.240\\\hline
F2   &  0.370 &  0.440 &  0.620 &  0.730 &  \textcolor{blue}{0.180} &  \textcolor{blue}{0.110} &  \textcolor{blue}{0.290}\\\hline
F5   &  0.470 &  0.570 &  0.790 &  0.910 &  \textcolor{blue}{0.220} &  \textcolor{blue}{0.120} &  \textcolor{blue}{0.340}\\\hline
\textcolor{red}{F8}&  \textcolor{red}{0.700} &  0.870 &  1.170 &  1.340 &  \textcolor{blue}{0.300} &  \textcolor{blue}{0.170} &  \textcolor{blue}{0.470}\\\hline
\textcolor{red}{G0}&  \textcolor{red}{0.900} &  \textcolor{red}{1.140} &  \textcolor{red}{1.520} &  \textcolor{red}{1.710} &  \textcolor{blue}{0.380} &  \textcolor{blue}{0.190} &  \textcolor{blue}{0.570}\\\hline
\textcolor{red}{G2}&  \textcolor{red}{1.060} &  \textcolor{red}{1.350} &  \textcolor{red}{1.800} &  \textcolor{red}{1.990} &  \textcolor{blue}{0.450} &  \textcolor{blue}{0.190} &  \textcolor{blue}{0.640}\\\hline
\textcolor{red}{G3}   &  \textcolor{red}{1.120} &  \textcolor{red}{1.430} &  \textcolor{red}{1.900} &  \textcolor{red}{2.090} &  \textcolor{blue}{0.470} &  \textcolor{blue}{0.190} &  \textcolor{blue}{0.660}\\\hline
G3.5 &  1.160 &  \textcolor{red}{1.470} &  \textcolor{red}{1.950} &  \textcolor{red}{2.150} &  \textcolor{blue}{0.480} &  \textcolor{blue}{0.200} &  \textcolor{blue}{0.680}\\\hline
G4   &  1.190 &  \textcolor{red}{1.520} &  \textcolor{red}{2.010} &  \textcolor{red}{2.200} &  \textcolor{blue}{0.490} &  \textcolor{blue}{0.190} &  \textcolor{blue}{0.680}\\\hline
G5   &  1.260 &  \textcolor{red}{1.610} &  2.130 &  \textcolor{red}{2.320} &  \textcolor{blue}{0.520} &  \textcolor{blue}{0.190} &  \textcolor{blue}{0.710}\\\hline
G8   &  1.430 &  1.830 &  2.410 &  2.590 &  \textcolor{blue}{0.580} &  \textcolor{blue}{0.180} &  \textcolor{blue}{0.760}\\\hline
K0   &  1.590 &  2.010 &  2.640 &  2.800 &  \textcolor{blue}{0.630} &  \textcolor{blue}{0.160} &  \textcolor{blue}{0.790}\\\hline
K1   &  1.680 &  2.110 &  2.760 &  2.910 &  \textcolor{blue}{0.650} &  \textcolor{blue}{0.150} &  \textcolor{blue}{0.800}\\\hline
K2   &  1.760 &  2.200 &  2.870 &  3.010 &  \textcolor{blue}{0.670} &  \textcolor{blue}{0.140} &  \textcolor{blue}{0.810}\\\hline
K3   &  1.960 &  2.410 &  3.140 &  3.250 &  \textcolor{blue}{0.730} &  \textcolor{blue}{0.110} &  \textcolor{blue}{0.840}\\\hline
K3.5 &  2.040 &  2.500 &  3.250 &  3.340 &  \textcolor{blue}{0.750} &  \textcolor{blue}{0.090} &  \textcolor{blue}{0.840}\\\hline
K4   &  2.130 &  2.590 &  3.370 &  3.440 &  \textcolor{blue}{0.780} &  \textcolor{blue}{0.070} &  \textcolor{blue}{0.850}\\\hline
K5   &  2.270 &  2.740 &  3.550 &  3.590 &  \textcolor{blue}{0.810} &  \textcolor{blue}{0.040} &  \textcolor{blue}{0.850}\\\hline\hline

Spectral Type & \multicolumn{7}{c|}{Intrinsic colors of MS Stars \citep{2001ApJ...558..309D}} \\\hline

A9   & 0.310 & 0.310 & 0.490 & 0.440 & \textcolor{blue}{0.180} & \textcolor{blue}{-0.050} & 0.130\\\hline
F0   & 0.360 & 0.370 & 0.570 & 0.520 & \textcolor{blue}{0.200} & \textcolor{blue}{-0.050} & 0.150\\\hline
F1   & 0.400 & 0.430 & 0.640 & 0.580 & \textcolor{blue}{0.210} & \textcolor{blue}{-0.060} & 0.150\\\hline
F2   & 0.450 & 0.480 & 0.710 & 0.660 & \textcolor{blue}{0.230} & \textcolor{blue}{-0.050} & 0.180\\\hline
\textcolor{red}{F5}   & \textcolor{red}{0.570} & 0.670 & 0.930 & 0.890 & \textcolor{blue}{0.260} & \textcolor{blue}{-0.040} & 0.220\\\hline
\textcolor{red}{F8}   & \textcolor{red}{0.640} & 0.790 & 1.060 & 1.030 & \textcolor{blue}{0.270} & \textcolor{blue}{-0.030} & 0.240\\\hline
\textcolor{red}{G0}   & \textcolor{red}{0.700} & 0.870 & 1.150 & 1.140 & \textcolor{blue}{0.280} & \textcolor{blue}{-0.010} & 0.270\\\hline
\textcolor{red}{G2}   & \textcolor{red}{0.750} & \textcolor{red}{0.970} & 1.250 & 1.260 & \textcolor{blue}{0.280} & \textcolor{blue}{0.010} & \textcolor{blue}{0.290}\\\hline
\textcolor{red}{G3}   & \textcolor{red}{0.760} & \textcolor{red}{0.980} & 1.270 & 1.280 & \textcolor{blue}{0.290} & \textcolor{blue}{0.010} & \textcolor{blue}{0.300}\\\hline
\textcolor{red}{G5}   & \textcolor{red}{0.780} & \textcolor{red}{1.020} & 1.310 & 1.320 & \textcolor{blue}{0.290} & \textcolor{blue}{0.010} & \textcolor{blue}{0.300}\\\hline
\textcolor{red}{G8}   & \textcolor{red}{0.850} & \textcolor{red}{1.140} & \textcolor{red}{1.440} & 1.470 & \textcolor{blue}{0.300} & \textcolor{blue}{0.030} & \textcolor{blue}{0.330}\\\hline
\textcolor{red}{K0}   & \textcolor{red}{0.970} & \textcolor{red}{1.340} & \textcolor{red}{1.670} & \textcolor{red}{1.740} & \textcolor{blue}{0.330} & \textcolor{blue}{0.070} & \textcolor{blue}{0.400}\\\hline
\textcolor{red}{K1}   & \textcolor{red}{1.050} & \textcolor{red}{1.460} & \textcolor{red}{1.800} & \textcolor{red}{1.890} & \textcolor{blue}{0.340} & \textcolor{blue}{0.090} & \textcolor{blue}{0.430}\\\hline
K2   & 1.140 & \textcolor{red}{1.600} & \textcolor{red}{1.940} & \textcolor{red}{2.060} & \textcolor{blue}{0.340} & \textcolor{blue}{0.120} & \textcolor{blue}{0.460}\\\hline
K3   & 1.250 & 1.730 & \textcolor{red}{2.090} & \textcolor{red}{2.230} & \textcolor{blue}{0.360} & \textcolor{blue}{0.140} & \textcolor{blue}{0.500}\\\hline
K4   & 1.340 & 1.840 & 2.220 & \textcolor{red}{2.380} & \textcolor{blue}{0.380} & \textcolor{blue}{0.160} & \textcolor{blue}{0.540}\\\hline
K5   & 1.540 & 2.040 & 2.460 & 2.660 & \textcolor{blue}{0.420} & \textcolor{blue}{0.200} & \textcolor{blue}{0.620}\\\hline
K7   & 1.860 & 2.300 & 2.780 & 3.010 & \textcolor{blue}{0.480} & \textcolor{blue}{0.230} & \textcolor{blue}{0.710}\\\hline
M0   & 2.150 & 2.490 & 3.040 & 3.290 & \textcolor{blue}{0.550} & \textcolor{blue}{0.250} & \textcolor{blue}{0.800}\\\hline
M1   & 2.360 & 2.610 & 3.220 & 3.470 & \textcolor{blue}{0.610} & \textcolor{blue}{0.250} & \textcolor{blue}{0.860}\\\hline
M2   & 2.620 & 2.740 & 3.420 & 3.670 & \textcolor{blue}{0.680} & \textcolor{blue}{0.250} & \textcolor{blue}{0.930}\\\hline
M3   & 2.840 & 2.840 & 3.580 & 3.830 & \textcolor{blue}{0.740} & \textcolor{blue}{0.250} & \textcolor{blue}{0.990}\\\hline
M4   & 3.070 & 2.930 & 3.740 & 3.980 & \textcolor{blue}{0.810} & \textcolor{blue}{0.240} & \textcolor{blue}{1.050}\\\hline

\hline
\end{tabular}
}
\end{center}
\tablecomments{Top: $V-I$, Visual-NIR, and NIR colors for the A HST star/Subaru NIR detections with their corresponding uncertainties. Middle/Bottom: The intrinsic colors of supergiants and main-sequence stars from \citet{2001ApJ...558..309D}. The $V-I$, Visual-NIR (red), and NIR (blue) color ranges of the A HST star/Subaru NIR detections are marked in the middle and bottom tables, indicating a potential spectral type range for each luminosity class.} 
\tablenotetext{a}{The $J-H$, $H-K_{S}$, and $J-K_{S}$ indices for the middle and bottom tables are calculated from the $-(V-J) + (V-H)$, $-(V-H) + (V-K_{S})$, and $-(V-J) + (V-K_{S})$, respectively.}
\end{table}

\subsection{An extra-galactic supergiant or a galactic dwarf?}
\label{giant_or_dwarf}
There are two possibilities for the nature of the A HST star. It can be either a distant supergiant belonging to the recently formed stars along the Cen A jet or a faint MS dwarf belonging to our own Milky Way galaxy. Following the intrinsic colors of supergiants and MS stars given by \citet{2001ApJ...558..309D} (Table \ref{tab:colors}), the $V-I = 0.84 \pm 0.28$ color of the A HST star is consistent with either an $F8-G3$ supergiant, or an $F5-K1$ MS dwarf. 

The primary criterion used to assign a spectral type to the A HST star is the $V-I$ color, because of its smaller uncertainty. As a secondary criterion we use the visual-NIR colors, in order to check for inconsistencies in their overlap with the $V-I$. Finally (and for completeness), we also examine the overlap of the NIR with the other colors for each luminosity class. The results are presented in Table \ref{tab:colors}. 

The $V-I$ and visual-NIR colors of the A HST star suggest limited spectral type ranges, which, generally, agree for both luminosity classes, although in the case of supergiants the color overlap is nearly perfect. On the other hand, the NIR colors suggest much larger spectral type ranges for both luminosity classes, but despite they clearly overlap with the $V-I$ and visual-NIR colors, they cannot be individually used to draw useful conclusions.

We note that we do not correct for reddening as its exact value on the lines-of-sight towards our objects cannot be precisely known. According to the IRSA on-line dust extinction tool\footnote{\url{http://irsa.ipac.caltech.edu/applications/DUST/}}, however, the extinction on large-- several arcmin-- scales towards NGC5128 should not be significant. Its effect on the colors of the A HST star would be a reduction of the $V-I$ by $\sim0.12$ mag, of the visual-NIR by $\sim0.20-0.24$ mag, and of the NIR colors by $\sim0.02-0.04$ mag. These extinction corrections are in excellent agreement with recent-- higher resolution-- results from \cite{2016A&A...595A..65S} for regions near our field. Accounting for these would shift our classifications towards $\sim1$ spectral type earlier in Table \ref{tab:colors}, which will not, qualitatively at least, affect our interpretation.\footnote{Visually, this shift would place our stars near the left tip of their horizontal error-bars in the visual-NIR CMD of figure \ref{fig:rgb_sequences}.}

So, is there a way to further distinguish between the extra-galactic supergiant and the galactic dwarf scenarios? 

\subsubsection{Supergiant}
\label{supergiant}

If the A HST star is an extra-galactic supergiant, then the remaining steps are relatively straightforward. Adopting a distance modulus for NGC5128 of $\mu = 27.92 \pm 0.19$ mag \citep{2004A&A...413..903R}, its absolute magnitude $M_{HST,A}^{V}=-7.69 \pm 0.22$ mag combined with its $V-I$ color range place the A HST star in the region of the Hertzsprung-Russell (HR) diagram occupied by YSGs (e.g a $G1$). In order to estimate its age, we use isochrones for rotating and non-rotating stars with $Z=Z{_\odot}$ from \citet{2012A&A...537A.146E} and isochrones for non-rotating stars with solar and sub-solar \citep[i.e. $Z\sim0.6Z{_\odot}$,][]{2001A&A...379..781R,2016A&A...586A..45S} metallicities from \citet{2009A&A...508..355B}. The best fitting isochrones indicate an age\footnote{The age is calculated as the median age of the within-the-photometric-uncertainties isochrones from all the models we used.} of $\sim10^{+4}_{-3}$ Myr for the A HST star. An example of the isochrones used is shown in the visual-NIR CMD of figure \ref{fig:rgb_sequences}. 

Finding young supergiants in this part of Cen A, however, should not be a surprise. Earlier HST \citep{2000ApJ...536..266M} and VLT observations \citep{2001A&A...379..781R} confirmed the presence of young massive stars in the region, likely to have formed during the most recent phases of the jet-HI cloud interaction that powers the SF of the region over the last $\sim100$ Myr \citep{2004A&A...415..915R}. In such a framework young, later-type supergiants are naturally expected in the vicinity, since the most massive ($\gtrsim15-20~M_{\sun}$) of the later ($\lesssim10-15$ Myr) generations of blue MS stars are expected to have evolved past the MS. 

\begin{table*}[t]
\caption{Parameters for the B and C HST stars.}
\label{tab:star}
\begin{center}
\begin{tabular}{|c|c|c|c|c|c|c|c|c|c|}\hline
\multicolumn{10}{|c|}{Extra-galactic Supergiant}\\\hline
 HST Star   & $V-I$ & \multicolumn{2}{c|}{S.T.R.} & I.S.T.\tablenotemark{a}  & $m_{V}$ & $M_{V}$ & Exp. $m_{K_{S}}$ & Exp. $m_{H}$ & Exp. $m_{J}$ \\\hline
units & mag & \multicolumn{2}{c|}{-} & - & mag & mag & mag & mag & mag \\\hline
uncert. & $\pm0.28$ & \multicolumn{2}{c|}{-} & - & $\pm0.12$ & $\pm0.22$ & - & - & - \\\hline
B (red) & 0.68 & \multicolumn{2}{c|}{$F5-G0$} & $F8$ & 21.24 & -6.68 & 19.90$_{-0.37}^{+0.43}$ & 20.07$_{-0.35}^{+0.38}$ & 20.37$_{-0.27}^{+0.30}$\\\hline
C (yellow) & 1.97 & \multicolumn{2}{c|}{$K2-K4$} & $K3/3.5$ & 23.20 & -4.72 & 19.91$_{-0.15}^{+0.28}$ & 20.01$_{-0.18}^{+0.32}$ & 20.75$_{-0.14}^{+0.25}$\\\hline\hline
 \multicolumn{10}{|c|}{Galactic Dwarf}\\\hline
HST Star   & $V-I$ & S.T.R. & I.S.T.\tablenotemark{a}  & $m_{V}$ & $\mu$ & \multicolumn{2}{c|}{Lum. Dist.} & \multicolumn{2}{c|}{Vert. Dist.}\\\hline
units & mag & - & - & mag & mag & \multicolumn{2}{c|}{kpc} & \multicolumn{2}{c|}{kpc}\\\hline
uncert. & $\pm0.28$ & - & - & $\pm0.12$ & - & \multicolumn{2}{c|}{-} & \multicolumn{2}{c|}{-}\\\hline
B (red) & 0.68 & $F1-G8$ & $G0$ & 21.24 & 16.54$_{-0.90}^{+1.50}$ & \multicolumn{2}{c|}{20.32$_{-6.89}^{+20.23}$} & \multicolumn{2}{c|}{6.82$_{-2.31}^{+6.78}$}\\\hline
C (yellow) & 1.97 & $K7-M0$ & $K7/M0$ & 23.20 & 14.80$_{-0.30}^{+0.30}$ & \multicolumn{2}{c|}{9.12$_{-1.18}^{+1.35}$} & \multicolumn{2}{c|}{3.06$_{-0.40}^{+0.45}$}\\\hline

\hline
\end{tabular}
\end{center}
\tablecomments{The first five columns are common in both tables. These indicate 1) the HST star identification corresponding to the letters/colored apertures of Fig. \ref{fig:SUB_HST_FOV}, 2) the $V-I$ color, 3) the spectral type range (S.T.R.) constrained through the suggested $V-I$ color on Table \ref{tab:colors}, 4) the intermediate spectral type (I.S.T.), and 5) the apparent $V$ band magnitude. Top table: For the extra-galactic supergiant scenario, column 6) shows the absolute $V$ band magnitude assuming a distance modulus of $\mu = 27.92 \pm 0.19$ mag for NGC5128, while columns 7), 8), and 9) show the expected $K_{S}$, $H$, and $J$ band apparent magnitudes, respectively, assuming $V-K_{S}$, $V-H$, and $V-J$ colors from Table \ref{tab:colors}, for the chosen I.S.T. Bottom table: For the galactic dwarf scenario, column 6) shows the estimated distance modulus for the chosen I.S.T., column 7) shows the estimated luminosity distance, and column 8) shows the estimated vertical to the galactic plane (luminosity) distance, using a galactic latitude of $\sim19.6^{\circ}$.}
\tablenotetext{a}{The I.S.T. is chosen to be the middle spectral type of the corresponding S.T.R., or the average of the middle spectral types in the case of an even S.T.R. If the S.T.R. consists only of two spectral types, then the I.S.T. is their average. The uncertainties in the columns 7), 8), and 9) of extra-galactic supergiants and 6), 7), and 8) of galactic dwarfs cover the entire corresponding S.T.Rs.}

\end{table*}

\subsubsection{Dwarf}
\label{dwarf}

If the A HST star is a foreground MS dwarf, then we need to constrain its distance in order to get an estimate on its location within our Galaxy. 

According to the suggested $V-I$ range (lower Table \ref{tab:colors}), its spectral type should be between $F5$ and $K1$. Assuming an intermediate spectral type to this range (e.g. $G3$), the expected absolute magnitude of the A HST star should be $M^{V}_{G3~dwarf} \sim 5.10^{+1.10}_{-1.40}$ mag. From this and its apparent magnitude follows its distance modulus $\mu\sim15.13^{+1.40}_{-1.10}$ mag and, therefore, its luminosity distance  $D^{G3~dwarf}_{L}\sim10.62^{+9.61}_{-4.22}$ kpc. Accounting also for its galactic latitude of $19.6^{\circ}$, it should be located at a vertical to the galactic plane distance of $D^{G3~dwarf}_{\perp}\sim3.56^{+3.23}_{-1.41}$ kpc. The latter suggests that if the A HST star is a galactic source, then it should be located inside the (low density) stellar halo of the Milky Way \citep[e.g.][]{1988ada..book.....V,1997ASSL..212..133N}. 

Adopting a galactic stellar halo density profile from \citet{2011MNRAS.416.2903D} and integrating over the solid angle defined by the HiCIAO FoV along the line-of-sight towards Cen A, we can estimate the theoretical fraction of galactic halo stellar mass that we expect to probe with our field. This corresponds to a total stellar mass of $M^{Total}_{FoV}\sim0.39^{+0.39}_{-0.23} M_{\odot}$\footnote{The upper and lower limits reflect the different mass-to-light ratios assumed for the stellar halo of the Milky Way by \citet{2011MNRAS.416.2903D}, i.e. $\sim1-5$.} and it could represent a $K0-M8$ MS dwarf, assuming that the total probed stellar mass present in our field is in the form of one halo star. This mass range, however, overlaps only marginally with the potential mass range of the A HST star, namely, $\sim0.7-1.3 M_{\odot}$\footnote{Corresponding to spectral types $\sim F5-K1$.}. Considering also a) that the IMF peaks at $\sim0.18-0.25 M_{\odot}$ and traces the present day mass function (PDFM) for low mass stars relatively accurately \citep[e.g.][]{2010AJ....139.2679B} and b) that the color overlap is not as consistent as in the case of supergiants, makes the galactic dwarf scenario significantly less likely for the A HST star, although it cannot be conclusively excluded.

\vspace{0.1cm}
\subsection{What about the second and the third brighter HST stars?}
\label{red_circle}

The analysis, so far, suggests that the $>3\sigma$ NIR signal that we detect comes from the NIR counterparts of the brighter star in the HST field, which, likely, is an extra-galactic YSG located in a region of recent jet-induced SF in the halo of NGC5128. We do not detect, however, neither the second (B HST star) nor the third (C HST star) brighter HST stars (Fig. \ref{fig:SUB_HST_FOV}) in the NIR. 

Similarly to Sect. \ref{giant_or_dwarf}, we attempt to constrain the spectral types of the B and C HST stars (using only their $V-I$ colors) for both luminosity classes and estimate the corresponding quantities for the extra-galactic supergiant and foreground dwarf scenarios. The results are summarized in Table \ref{tab:star}. 

If these are extra-galactic stars, then their HST V band photometry (upper Table \ref{tab:star}) and the distance modulus of NGC5128 suggest absolute visual magnitudes that place them in the supergiant region of the HR diagram. According to these and their $V-I$ color ranges, the B HST star is a YSG (i.e. an $F8$), while the C HST star is an RSG (i.e. a $K3/3.5$). Similarly to Sect. \ref{supergiant}, we use various isochrones in order to estimate the ages of the B and C HST stars, namely, $\sim16^{+6}_{-3}$ Myr and $\sim25^{+15}_{-9}$ Myr, respectively. 

In the supergiant scenario, therefore, both of these stars appear to be older than the A HST star. Despite the uncertainties in our age estimations this could be viewed as an additional indication that the recent SF in the halo of NGC5128 is not the result of a single episodic SF event, but rather of a continuous process \citep{2004A&A...415..915R}. 

If these stars are foreground MS dwarfs (lower Table \ref{tab:star}), then their expected vertical to the galactic plane distances would place them inside the stellar halo of the Milky Way. Similarly to Sect. \ref{dwarf}, their spectral type ranges suggest potential masses of $\sim0.8-1.5 M_{\odot}$ for the B and $\sim0.5 M_{\odot}$ for the C HST stars. Evidently, the total expected stellar halo mass probed by the HiCIAO FoV overlaps with the expected mass of the C HST star, but not with that of the B HST star. This makes the galactic dwarf scenario potentially plausible only for the C HST star, leaving the B HST star as a good candidate supergiant in Cen A.

As noted earlier (Sect. \ref{dwarf}), however, the C HST star can be considered as a foreground star under the assumption that all the expected halo stellar mass within our field is in the form of one halo star. Under the more conservative assumption that the probed halo stellar mass is in the form of the sum of a few lower mass stars near the peak of the IMF/PDMF (i.e. $\sim M5-M8$), we consider as reasonable to treat the C HST star as a supergiant in Cen A for the rest of this work, since accounting for either two, or three candidate supergiants will not alter our conclusions.

This view is also supported by the work of \cite{2001A&A...379..781R}. They corrected their $6.8 \times 6.8$ arcmin FORS1 data of the region for foreground contamination using a Besancon group stellar population synthesis model \citep{Robin1986, Robin1996} of the Milky Way and they estimate the number of foreground stars to be $\sim340$\footnote{For a photometric depth of $V=25$ mag. This number should be, at most, an upper limit for our case, since the HST data used here have a photometric depth of $\sim24$ mag \citep{2000ApJ...536..266M}.}. Taking into account that the foreground stars are distributed homogeneously across their FoV (see their Fig. 15) and scaling the estimated number with the area, we expect $\lesssim0.8$ foreground stars in our $20 \times 20$ arcsec FoV, which suggests that all three HST stars are later-type supergiants in Cen A, in very good agreement with our earlier estimations.

\section{Discussion}
\label{sec:discussion}

We have extensively studied high resolution Subaru $J$, $H$, and $K_{S}$ and archived HST $F555W$ and $F814W$ imaging data of a region along the Cen A's jet path, in which jet-induced SF is thought to be taking place. 

We show that the NIR signal that we detect is statistically significant and that it originates from the position of the A HST star, the brighter field point source detected on the HST imaging data. Its characteristics are constrained using its $V-I$, visual-NIR, and NIR color indices, which suggest that, most likely, this star is a YSG located in Cen A. Similarly, but based only on their HST photometry (i.e. on the $V-I$ color), the extra-galactic supergiant scenario is promoted also for the B (YSG) and C (RSG) HST stars.

\begin{figure*}[t]
\centering
\includegraphics[width=15cm]{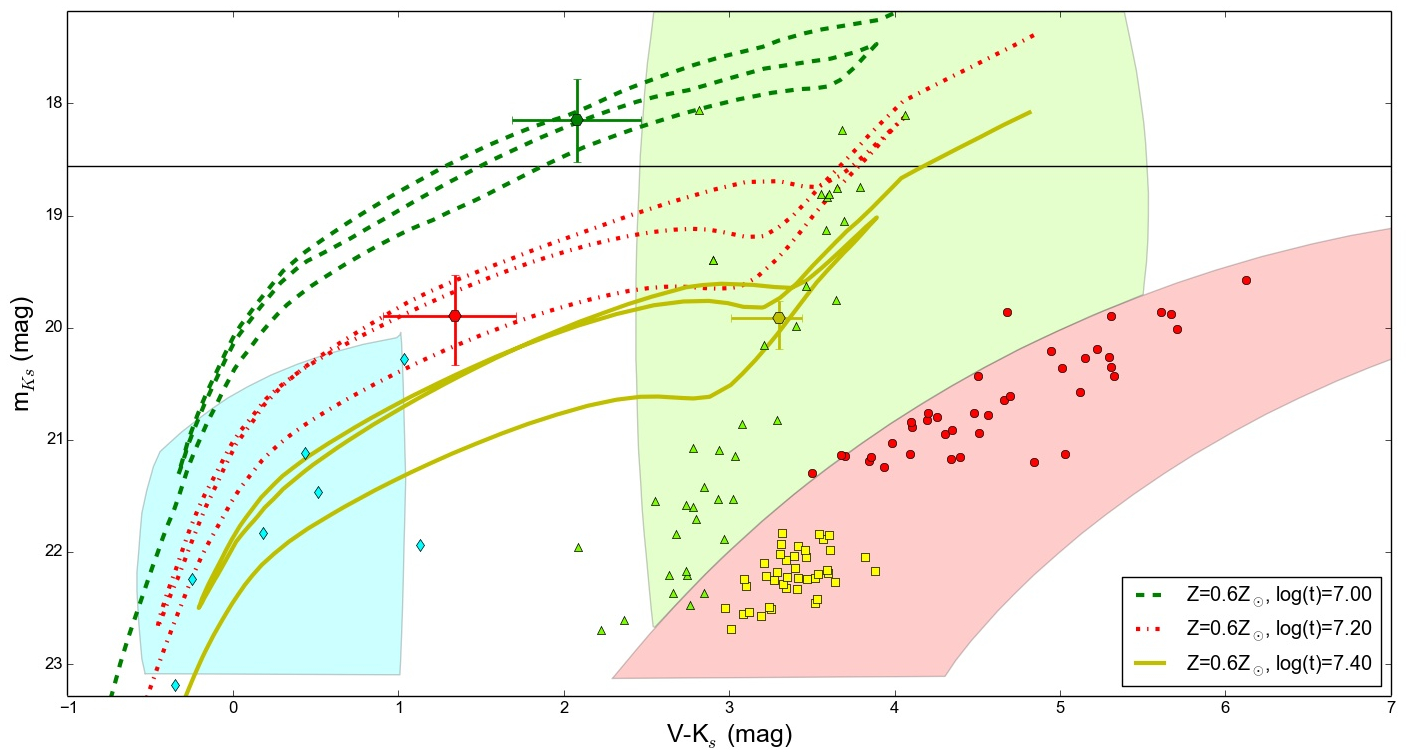}
\caption{Visual-NIR CMD of Fig. 3 from \citet{2008A&A...483L...5G}. Plotted are the BSGs (diamond symbols), RSGs (triangles), RGB (squares), and AGB (filled circles) stars of UKS 2323-326. The red, green, and blue shaded areas indicate the respective stellar sequences of the NE part of NGC5128's halo, estimated from Fig. 12 of \citet{2001A&A...379..781R}. The green, red, and yellow points with the error bars indicate the positions of the A, B, and C HST stars, respectively, on the CMD and their corresponding uncertainties. The dashed (green), the dashed-dotted (red), and the solid (yellow) lines are the isochrones from \citet{2009A&A...508..355B} for the median ages of the A, B, and C HST stars, respectively. Finally, the black solid line indicates the Subaru $K_{S}$ limiting magnitude.}
\label{fig:rgb_sequences}
\end{figure*}

\subsection{Evolved massive stars in Cen A}
\label{evolved massive stars}

The presence of later-type supergiants should be expected in this part of Cen A. Earlier studies \citep[e.g.][]{2000ApJ...538..594F,2000ApJ...536..266M,2001A&A...379..781R} have found young massive stars in this region, likely to have formed during the later ($\lesssim10-15$ Myr) phases of the jet-HI cloud interaction that is thought to drive the local SF over the last $\sim100$ Myr \citep{2004A&A...415..915R}. Our age estimates for the A, B, and C HST stars ($\sim10^{+4}_{-3}$, $\sim16^{+6}_{-3}$, and $\sim25^{+15}_{-9}$ Myr, respectively) appear to be consistent with a continuous SF process, although their uncertainties do not allow us to conclusively confirm it. Nonetheless, supergiants of different ages would naturally fit in such a framework, since a $\sim100$ Myr period of jet-HI gas interaction is expected to have created multiple generations of blue massive stars, each one evolving from the MS to the supergiant phase at a different (mainly mass-dependent) pace. 

Future, deep, spectro-photometric, optical/NIR AO observations of the entire region are expected to put a much better constraint on the ages of these evolved massive stars, which, along with their spatial distribution, would help us ``reverse engineer'' the most recent history of the jet-HI cloud interaction in Cen A, by investigating for example whether the jet is precessing, or not \citep[e.g.][]{2002ApJ...564..688R}. Since this is our closest known paradigm of such an interaction, understanding the properties of the evolved massive stellar populations of the region is essential for a deeper understanding of the underlying physics of positive AGN feedback.

Finally, finding candidate YSGs is also important for developing the stellar evolutionary theory, since the post MS evolution of massive stars is still uncertain \citep[e.g.][]{2012ARA&A..50..107L}. YSGs are located in the instability strip ($T_{eff}\sim 5000-6000$ K) at the upper-part of the HR diagram in the so-called ``HR gap'' \citep[e.g][]{2014A&A...570L..13C} and they can be either evolving from the MS to later spectral types, or from the RSG phase to earlier spectral types \citep[e.g][]{2016ApJ...825...50G}. Both of these transitions are predicted to occur very quickly during the post MS phase, which can explain the small number of stars detected in the HR gap \citep[e.g][]{2012ARA&A..50..107L}. In this sense, each candidate YSG detected is important, since it can provide further empirical constraints to the stellar evolutionary theory.

\subsection{Resolving extra-galactic stars beyond our LG}
\label{resolve extra galactic stars}

Our attempt to resolve individual stars in galaxies outside our LG, with 8-m class telescopes observations, is not the first. Various authors \citep[e.g.][]{2001ApJ...548L.159B,2002ApJ...567..277B,2008A&A...485...41C,2013ApJ...767L..29O} have conducted spectroscopic studies of isolated O and B stars in several nearby galaxies, while \citet{2001A&A...379..781R} photometrically studied two fields in the stellar halo of NGC5128 (including our field) using FORS1/ISAAC VLT observations. These studies, however, do not use AO assisted observations. 

The only AO assisted study of extra-galactic stellar populations to our knowledge was performed by \citet{2008A&A...483L...5G}, using MAD/VLT observations. In their study of the central region of the dwarf irregular galaxy UKS 2323-326, they resolved blue supergiants (BSGs), RSGs, red giant branch (RGB), and AGB stars down to $\sim 22$ mag in the $K_{S}$ band. This study is the most relevant to ours, since we both used AO assisted ground-based observations in order to resolve denser-than-average extra-galactic fields.

A direct comparison between the stellar populations of NGC5128 and UKS 2323-326 is shown in Fig. \ref{fig:rgb_sequences}. In this plot we show the CMD of Fig. 3 from \citet{2008A&A...483L...5G}, with our A, B, and C HST stars over-plotted. We also plot the-- roughly outlined-- sequences of BSGs, RSGs, and RGB/AGB stars\footnote{We choose the outline of the main distribution of the stars of Fig. 12 from \citet{2001A&A...379..781R} as the AGB/RGB branch of NGC5128, while we distinguish between its BSGs and RSGs by using approximate $V-K_{S}$ cut-offs from Table \ref{tab:colors}, namely $V-K_{S}<1$ and $2.5<V-K_{S}<5.5$, respectively.} of the NE part of NGC5128's halo, taken from the visual-NIR CMD of \citet{2001A&A...379..781R}. For this, we project the stars of UKS 2323-326 to the distance of NGC5128, by adding the difference of the distance moduli of the two galaxies, namely $\delta \mu = 1.18\pm0.24$\footnote{The adopted distance modulus of UKS 2323-326 is $\mu^{UKS 2323-326}=26.74\pm0.15$ mag \citep{2008A&A...483L...5G}.}, to all apparent stellar magnitudes of UKS 2323-326. Finally, the limiting magnitude of our $K_{S}$ band is indicated by the black solid line.

Although the stellar populations of NGC5128 seem to overlap with those of UKS 2323-326 quite well, they extend towards redder $V-K_{S}$ values, covering a wider area on the CMD and leading to an offset between their center lines. This could be the effect of the wider range of metallicities in the stars of NGC5128 \citep{2001A&A...379..781R}, a view consistent with the fact that, generally, dwarf galaxies have significantly lower metallicities than giant galaxies \citep{1998ARA&A..36..435M}. Moreover, this offset appears to be more dominant in the redder populations, since the BSGs of the two galaxies appear to overlap more consistently than the RSGs and the RGB/AGB stars do, implying that this offset is not a systematic error. This is an additional indication that the position of the A HST star in the YSG part of the CMD is relatively accurately determined. Other factors could also contribute to this offset, such as uncertainties in the distance moduli and/or differences in the reddening between the two fields, but these are not expected to significantly alter the global picture. 

Observationally, however, our task was somewhat more difficult. On the one hand, NGC5128 is by a factor of $\sim1.5-2$ more distant than UKS 2323-326, while on the other, the observations of \citet{2008A&A...483L...5G} were by a factor of $\sim3-4$ deeper\footnote{In terms of integration time. This reflects a better SNR by a factor of $\sim 5$ with respect to our Subaru data.} than our Subaru data. This is easily visible on the CMD of Fig. \ref{fig:rgb_sequences}. The position of the black solid line indicates that, at best, our observations would enable us to probe only a handful of stars on the brightest tip of NGC5128's YSG/RSG branch in the much larger field studied by \citet{2001A&A...379..781R}. The fact that we detect one of these stars is a consequence of the chosen field in a region of recent SF, which by itself increases the probability for such an observation. 

Our analysis, therefore, fully supports the views of \citet{2008A&A...483L...5G}. It suggests that it is now possible to resolve extra-galactic denser-than-average fields potentially into individual stars in galaxies located beyond our LG, using sufficiently deep AO assisted data from the current 8/10-m class telescopes. Further development of the AO systems and the upcoming class of ELT telescopes are expected to successfully undertake significantly more difficult tasks, pushing our understanding of the properties of the stellar populations of distant galaxies to a new level.

\section{Summary}
\label{sec:summary}

We have performed Subaru NIR and archived HST data analysis of a field in a region of recent SF along the Cen A's jet path. The key points of our analysis are: 
\begin{enumerate}

\item{In all NIR low-pass filtered images we see strong ($>3\sigma$ above the background mean) signal at the expected position of the brightest star in the equivalent HST field. The probability that this signal results from stochastic background fluctuations alone, at the same position, in three independent measurements is negligible ($\leq2.5\times10^{-7}\%$), implying that the NIR signal originates from the NIR counterparts of this star.}

\item{The very good overlap of the visual and NIR colors of the brightest HST star and its absolute magnitude suggest that it is an extra-galactic YSG with an estimated age of $\sim10^{+4}_{-3}$ Myr. The age of this star is consistent with the ages of the young blue stars that were previously identified in the region, thought to have formed after a recent jet-HI cloud interaction.}

\item{Based solely on their HST photometry, the second (YSG) and the third (RSG) brighter HST stars are, likely, also later-type supergiants in NGC5128 with estimated ages of $\sim16^{+6}_{-3}$ Myr and $\sim25^{+15}_{-9}$ Myr, respectively. The ages of the three supergiants and their indicated spread appear to be consistent with the view that the jet-HI cloud interaction was not a single episodic event, but, in fact, a continuous process during the past $\sim100$ Myr. Under certain conditions, however, the third brighter HST star could also be a foreground galactic dwarf.}

\item{With deeper, optical/NIR, spectro-photometric observations of the entire region, using the state-of-the-art 8/10-m class telescopes equipped with AO systems, we can study a great variety of physical phenomena simultaneously and in great detail. By better determining the ages of the evolved massive stars and by accurately mapping their positions, we can constrain the most recent jet-HI cloud interaction(s) and, therefore, the underlying physics of positive AGN feedback, enrich the sample of YSGs and provide stellar parameters to the stellar evolutionary models community, and further constrain the more massive part of the IMF in this region. }

\end{enumerate}

\section*{Acknowledgements}

We would like to thank Ass. Prof. Elias Tsakas and M.Sc. Christos Christou for the useful discussions on the mathematical rigidity of the present work. The Subaru telescope operation team as well as the HiCIAO instrument team for their generous support. This work was supported by the Max Planck Society and the University of Cologne through the International Max Planck Research School (IMPRS) for Astronomy and Astrophysics as well as in part by the Deutsche Forschungsgemeinschaft (DFG) via grant SFB $956$. We had fruitful discussions with members of the European Union funded COST Action MP$0905$: Black Holes in a violent Universe and the COST Action MP$1104$: Polarization as a tool to study the Solar System and beyond.

\bibliographystyle{aasjournal}
\bibliography{CenA} 
\label{lastpage}
\end{document}